\documentstyle[aasms4,12pt]{article}
\begin{document}

\title{SUPERWIND MODEL OF EXTENDED LYMAN$\alpha$ EMITTERS
       AT HIGH REDSHIFT}

\author{Yoshiaki Taniguchi, \& Yasuhiro Shioya}

\affil{Astronomical Institute, Graduate School of Science, 
       Tohoku University, Aramaki, Aoba, Sendai 980-8578, Japan}

\begin{abstract}
We propose a new model for the extended Ly$\alpha$
blobs found recently at high redshift $z \sim$ 3.
The observational properties of these blobs are;
1) the observed Ly$\alpha$ luminosities are $\sim 10^{43} h^{-2}$ 
ergs s$^{-1}$, 2) they appear elongated morphologically,
3) their sizes amount to $\sim$ 100 kpc,
4) the observed line widths amount to $\sim 1000$ km s$^{-1}$,
and 5) they are not associated with strong radio-continuum sources.
All these observational properties seem to be explained in 
terms of galactic winds driven by successive supernova
explosions shortly after the initial burst of massive star 
formation in the galactic centers.
The observed number density of Ly$\alpha$ blobs 
($\sim 3.4 \times 10^{-5} h^3$ Mpc$^{-3}$) may be explained if
their present-day counterparts are elliptical galaxies
with a luminosity above $\sim 1 L^*$.
\end{abstract} 


\section{INTRODUCTION}

\subsection{Surveys for High-Redshift Galaxies}

Recent great progress in the observational astronomy
have revealed that a large number of high-redshift galaxies
can be accessible by continuum emission of galaxies
(stellar continuum, thermal continuum from dust grains, 
or nonthermal continuum from plasma heated by supernovae)
in a wide range of observed wavelengths between optical and radio
(e.g., Williams et al. 1996; Lanzetta, Yahil, \& Fern\'andez-Soto
1996; Chen, Lanzetta, \& Pascarelle 1999; Steidel et al. 1996a, 1996b;
Dey et al. 1998; Spinrad et al. 1998; Weymann et al. 1998;
van Breugel et al. 1999;
Smail et al. 1997; Hughes et al. 1998; Barger et al. 1998;
Eales et al. 1999; Barger, Cowie, \& Sanders 1999; Richards et al. 1999).
On the other hand, it has been also argued often that 
forming galaxies at high redshift experienced very luminous 
starbursts and thus they could be much brighter in line emission such as 
Ly$\alpha$ and [O {\sc ii}]$\lambda$3727 emission lines
(e.g., Partridge \& Peebles 1967; Larson 1974; Meier 1976). 
However, although many attempts have been made to search for such 
very strong emission-line sources at high redshift
(see for a review, Pritchet 1994; see also Pahre \& Djorgovski 1995; 
Thompson, Mannucci, \& Beckwith 1996),
most these searches failed except some successful
surveys around known high-$z$ objects such as quasars
(Hu \& McMahon 1996; Hu, McMahon,
\& Egami 1996; Petitjean et al. 1996; Hu, McMahon, \& Cowie 1999).
Very recently, 
a new attempt with the Keck 10 m telescope has revealed the 
presence of Ly$\alpha$ emitters in blank fields at high
redshift (Cowie \& Hu 1998, hereafter CH98).
Subsequently, Keel et al. (1999, hereafter K99) and Steidel
et al. (1999, hereafter S99) also found a number of 
high-$z$  Ly$\alpha$ emitters in other sky areas.
Since these three surveys have reinforced the potential importance of 
search for high-$z$ Ly$\alpha$ emitters,
it seems urgent to investigate the origin of Ly$\alpha$ emitters. 

\subsection{Extended Lyman $\alpha$ Emitters at High Redshift}

A brief summary of the recent three surveys  
for high-$z$ Ly$\alpha$ emitters (CH98, K99, and S99)
is given in Table 1. 
In this Letter, we adopt an Einstein-de Sitter cosmology with
a Hubble constant $H_0 = 100 h$ km s$^{-1}$ Mpc$^{-1}$.
Each survey has
discovered more than ten strong Ly$\alpha$ emitters with the Ly$\alpha$
equivalent width above 100 \AA~  in the observed frame.
It is interesting to note that five sources among them  are
observed to be very extended spatially, e.g., $\sim$ 100 kpc
(K99; S99); we call them Ly$\alpha$ blobs following S99
(hereafter LABs). These five sources are cataloged as
Object 18, Object 19, 53W002 (K99), Blob 1, and Blob 2 (S99).
Their basic data are given in Table 2.
Since the three LABs found in K99 are all strong C {\sc iv} emitters
(Pascarelle et al. 1996), it seems natural to conclude 
that they are photoionized by the central engine of active galactic
nuclei (AGNs) (K99).
On the other hand, the remaining two LABs found by S99 have no evidence 
for the association with AGNs (S99). It should be also noted that
their observed Ly$\alpha$ equivalent widths, $EW({\rm Ly}\alpha)$ 
$\sim$ 1500 \AA, are much larger than those of the three K99 sources.
These suggest that the origin of LABs may be heterogeneous and thus
the origin of S99 LABs is different from that of K99 ones. 

Here we summarize the observational properties of the LABs found by S99;
1) the observed Ly$\alpha$ luminosities are $\sim 10^{43} h^{-2}$
ergs s$^{-1}$, 2) they appear elongated morphologically,
3) their sizes amount to $\sim$ 100 $h^{-1}$ kpc,
4) the observed line widths amount to $\sim 1000$ km s$^{-1}$, and
5) they are not associated with strong radio-continuum sources
such as powerful radio galaxies. 
One possible origin may be that these LABs are 
superwinds driven by the initial starburst in galaxies because
i) a superwind could develop to a distance of $\sim$ 100 kpc
in the low-density intergalactic medium (IGM), and ii) a superwind often
blows with a bi-conical morphology (e.g., Heckman, Armus, \& Miley 1990).
In this Letter, we investigate this possibility. 
We also discuss a possible evolutionary link between LABs
and high-$z$, dust-enshrouded submm sources 
(Barger et al. 1999 and references therein).

\section{SUPERWIND MODEL}

\subsection{Superwinds From Forming Galaxies}

We consider the possibility that a LAB is 
a well-developed superwind seen from a nearly edge-on view. 
First, we investigate properties of 
a superwind caused by the initial starburst in a galaxy.
We adopt the dissipative collapse scenario for formation of elliptical 
galaxies and bulges [i.e., the  monolithic collapse
model\footnote{It is not necessarily to presume
that this pregalactic cloud is a first-generation gigantic gas cloud.
If a number of subgalactic gas clouds are assembled into one and then
a starburst occurs in its central region, the physical situation
seems to be nearly the same as that of the monolithic collapse.}
(Larson 1974)] together with the galactic
wind model proposed by  Arimoto \& Yoshii (1987; hereafter AY87;
see also Kodama \& Arimoto 1997). In this scenario,
the initial starburst occurs at the epoch of galaxy formation 
in the galaxy center.
Subsequently, massive stars die and then a large number of
supernovae appear. These supernovae could overlap and then evolve
into a so-called superbubble. If the kinetic energy deposited to the
surrounding gas overcomes the gravitational potential energy of the
galaxy, the gas clouds are blown out into the intergalactic space
as a superwind (e.g., Heckman et al. 1990).

The evolution of such a superwind can be
described  by superbubble models (McCray \& Snow 1979; Koo \& McKee
1992a, 1992b; Heckman et al. 1996; Shull 1995).
The radius and velocity of the shocked shells\footnote{It is
noted that the derivation of
$r_{\rm shell}$ requires that the baryonic component dominates
the gravitational potential. Although the presence of a dark matter
halo requires that this estimate of $r_{\rm shell}$
is not valid at arbitrarily large radii, we do not take account of
this effect because our discussion is an order-of-magnitude one.}
at time $t$ (in units of $10^8$ years) are then

\begin{equation}
r_{\rm shell}  \sim 110 
L_{\rm mech, 43}^{1/5}
n_{\rm H, -5}^{-1/5}
t_{8}^{3/5} ~~ {\rm kpc},
\end{equation}
and

\begin{equation}
v_{\rm shell} \sim 650 
L_{\rm mech, 43}^{1/5}
n_{\rm H, -5}^{-1/5}
t_{8}^{-2/5} {\rm km ~ s}^{-1},
\end{equation}
where $L_{\rm mech}$ is the mechanical luminosity 
released collectively from the supernovae in the central starburst
in units of $10^{43}$ ergs s$^{-1}$ and $n_{\rm H}$ is the average 
hydrogen number density of the IGM in units of $10^{-5}$ cm$^{-3}$.

We can estimate $L_{\rm mech}$ directly from AY87.
For an elliptical galaxy with a stellar mass 
$M_{\rm stars} = 10^{11} M_\odot$, radius  $r \simeq$ 10 kpc and
${n}_{\rm H} \sim 1$   cm$^{-3}$ (Saito 1979; AY87),
we expect $N_{\rm SN} \sim 3 \times 10^9$ stars that explode as supernovae.
Since most of these massive stars were formed during the first 
$5 \times 10^8$ years (= $t_{\rm GW}$), we obtain
$L_{\rm mech} \sim \eta ~ E_{\rm SN} ~ N_{\rm SN} / t_{\rm GW} \sim 10^{43} ~ 
{\rm erg~ s}^{-1}$ where $E_{\rm SN}$ is the
total energy of a single supernova ($10^{51}$ ergs) and
$\eta$ is the efficiency of the kinetic energy deposited to the ambient gas 
($\sim$ 0.1; Dyson \& Williams 1980).
We assume for simplicity that a hydrogen number density in the IGM
is $n_{\rm IGM}(z) \sim 0.1 n_{\rm cr}(z) =
0.1 n_{\rm cr}(0) (1+z)^3 \simeq 1.1 \times 10^{-6} h^{-2} (1+z)^3$
where $n_{\rm cr}(0)$
is the critical number density corresponding to the critical mass density
of the universe, $\rho_{\rm cr}(0) = 3 H_0^2 / (8 \pi G) \simeq
1.9 \times 10^{-29} h^2$ g cm$^{-3}$.
We thus obtain $n_{\rm IGM}(3) \simeq 7.3 \times 10^{-5} h^{-2}$ cm$^{-3}$
at $z = 3$.
Since we assume that the superwind is seen from
a nearly edge-on view, we obtain a characteristic size of the superwind,
$l \sim 2 \times r_{\rm shell} \sim$ 150 kpc with $n_{\rm H, -5} = 7.3$. 
If we assume an opening angle of the superwind $\theta_{\rm open} = 45^\circ$
(see section 2.3),
we obtain a full width at half maximum velocity of the superwind,
FWHM $\sim 2 \times v_{\rm shell} \sin \theta_{\rm open}
\simeq 620$ km s$^{-1}$. These values appear consistent with the 
observations (S99). 

\subsection{Frequency of Occurrence of Superwinds at High Redshift}

Since our superwinds model implies that the most probable progenitors 
of LABs are forming elliptical galaxies,
it is important to compare the observed
number density of LABs at high redshift with that of 
elliptical galaxies in the local universe. 
The observed number density of LABs at high redshift can be 
related to the number density of elliptical galaxies
responsible for the LABs $n_{\rm E-LAB}$ as

\begin{equation}
n_{\rm LAB} \sim n_{\rm E-LAB} \nu_{\rm SW} (1 - \Delta\Omega/4\pi),
\end{equation}
where $\nu_{\rm SW}$ is the chance probability to find 
superwinds from high-$z$ ellipticals,
and $\Delta\Omega$ is the full opening solid angle of a pair of 
the superwind in units of steradian. The last term is attributed to
the assumption that we observe superwinds from a nearly edge-on view.

First, based on the results of CH98, K99, and S99, we estimate $n_{\rm LAB}$
using the following relation,

\begin{equation}
n_{\rm LAB} =  {{N_{\rm LAB}({\rm CH98}) + N_{\rm LAB}({\rm K99}) +
N_{\rm LAB}({\rm S99})} \over {V({\rm CH98}) f_{\rm cl}({\rm CH98}) +
V({\rm K99}) f_{\rm cl}({\rm K99}) + V({\rm S99}) f_{\rm cl}({\rm S99})}}
\end{equation}
where $V$ is the co-moving volume of the surveyed area (see Table 1) and 
$f_{\rm cl}$ is the clustering factor
of galaxies in the surveyed volume with respect to the so-called field.
In the CH survey, no LAB is found in the two blank fields; i.e.,
$N_{\rm LAB}({\rm CH98}) =0$ and $f_{\rm cl}({\rm CH98}) = 1$.
In the S99 survey, the two LABs are found in the proto-cluster region
in which the number density of galaxies is higher by a factor of
$\approx$ 6 than that in the field; i.e., 
$N_{\rm LAB}({\rm S99}) =2$ and $f_{\rm cl}({\rm S99}) = 6$.
In the K99 survey, although the three LABs are found in the
53W002 field, all of them are associated with AGNs.
Therefore, we adopt $N_{\rm LAB}({\rm K99}) = 0$. There is a rich group
of galaxies in this field (Pascarelle et al. 1996). However, since it is 
difficult to estimate its clumping factor quantitatively, we assume 
$f_{\rm cl}({\rm K99}) \simeq  f_{\rm cl}({\rm S99}) = 6$.
We do not use the data of the other five fields surveyed by K99
because the detection limits of Ly$\alpha$ emission are higher 
by a factor of 2 than those of the other survey fields.
Then we obtain $n_{\rm LAB} \simeq 3.4 \times 10^{-5} h^{3}$ Mpc$^{-3}$.

Next we estimate the probability to observe superwinds $\nu_{\rm SW}$.
Since galaxies beyond $z \sim 5$ have been found (e.g., Hu et al. 1999;
Dey et al. 1998; Spinrad et al. 1998; Weymann et al. 1998;
van Breugel et al. 1999), we assume that elliptical galaxies
were formed randomly
at a redshift range between $z = 10$ and $z = 3$.
According to the cosmology model adopted here, the above redshift
interval corresponds to a duration of $\tau_{\rm form} \approx 
6.4 \times 10^8 ~ h^{-1}$ years. For an elliptical galaxy with 
a mass of $10^{11} M_\odot$, the galactic wind breaks at
$t({\rm SW}) \simeq 3.5 \times
10^8$ years after the onset of the initial starburst (AY87).
Therefore, superwinds could be observed from such ellipticals
with $z \lesssim 6$. The chance probability of superwinds 
can be estimated as $\nu_{\rm SW} = \tau_{\rm SW}/\tau_{\rm form}$
where $\tau_{\rm SW}$ is the duration when a superwind can be observed
as an emission-line nebula.
As shown in section 2.1, a duration of $\tau_{\rm SW} 
\approx 1 \times 10^8$ years is necessary to develop the superwind 
to a radius of $\sim 100 h^{-1}$ kpc.
Therefore, we obtain $\nu_{\rm SW} \simeq 0.16 h$.

Thirdly, we estimate the probability to observe superwinds from
a nearly edge-on view.
A typical semi-opening angle of superwinds may be
$\theta_{\rm open} \simeq 45^\circ$
(e.g., Heckman et al. 1990; Ohyama, Taniguchi, \& Terlevich 1997
and references therein). This gives
$1 -\Delta\Omega/4\pi = \cos \theta_{\rm open} \simeq 0.71$.

Then, we obtain

\begin{equation}
n_{\rm E-LAB} \sim n_{\rm LAB} \nu_{\rm SW}^{-1} (1 - \Delta\Omega/4\pi)^{-1}
\sim 3.0 \times 10^{-4} ~ h^2 ~ {\rm Mpc}^{-3}.
\end{equation}
Integrating the luminosity function of elliptical galaxies 
derived by Marzke et al. (1994), we find that the above 
number density corresponds to that of ellipticals above
$\simeq 1 L^*$ when $h$ lies in a range between 0.5 and 1. 
Since the mass of an elliptical galaxy
with $L_*$ is $\sim 10^{11} M_\odot$ (AY87; Kodama \& Arimoto 1997), 
our superwind model appears consistent with the observations.

\subsection{Obscured Host Galaxies}

Finally we give comments on the visibility of 
galaxies hosting superwinds.  
In our superwind model, the central starburst region
may be obscured by the surrounding gas and dust. 
Although AY87 assume that the superwind blows isotropically for
simplicity, actual superwinds tend to have a bi-conical morphology.
This implies that a lot of gas and dust
may be located in the host galaxy with a disk-like configuration,
being responsible for the collimation of superwinds.
These gas clouds are expected to absorb the radiation 
from the central star cluster if we observe superwinds 
from a nearly edge-on view. Let us consider a case that 
gas clouds with a total  mass of $M_{\rm gas}$
are uniformly distributed in a disk with a radius of $r$ and a full 
height of $d$. We estimate the average number density of gas
$n_{\rm H} = M_{\rm gas}/[\pi r^2 d m_{\rm H}]
\simeq 14 M_{\rm gas, 10} r_{10}^{-2} d_1^{-1}$ cm$^{-3}$
where $M_{\rm gas, 10}$ is in units of $10^{10} M_\odot$,
$r_{10}$ is in units of 10 kpc, $d_{1}$ is in units of 1 kpc,
and $m_{\rm H}$ is the mass of a hydrogen atom.
This gives an H {\sc i} column density
$N_{\rm H} = n_{\rm H} r \simeq 4.2 \times 10^{23}
M_{\rm gas, 10} r_{10}^{-1} d_1^{-1}$ atoms cm$^{-2}$
for an edge-on view toward the gas disk, corresponding to 
the visual extinction of $A_V \sim 280$ mag where we use a relation of
$A_V ({\rm mag}) = N_{\rm H}/(1.54 \times 10^{21} ~ {\rm cm}^{-2})$
(e.g., Black 1987).
Even if the gas-to-dust mass ratio is ten times smaller than that of 
our galaxy, the visual extinction is still large, $A_V \sim 30$ mag.
This may be responsible for the observed shortage of the ultraviolet
luminosities accounting for the Ly$\alpha$ line luminosities (S99).

The obscuration described above may also be responsible for
the observed large equivalent widths of the Ly$\alpha$ emission
in S99; i.e., $EW({\rm Ly}\alpha) \sim 1500$ \AA. 
Since these LABs are observed at $z \approx 3.1$ (see Table 2),
the rest-frame equivalent widths are estimated to be 
$EW^0({\rm Ly}\alpha) \sim 375$ \AA. 
This value is still larger by a factor of two than those expected for
star-forming, dust-free galaxies; e.g., $EW({\rm Ly}\alpha) 
\simeq$ 50 -- 200 \AA~ (Charlot \& Fall 1993; see also Tenorio-Tagle
et al. 1999). 
However, in our model, strong continuum radiation from
the central star cluster can be obscured by a lot of surrounding
gas and dust. On the other hand, the Ly$\alpha$ emission arises 
from the superwind which is far from the host, e.g., $r \sim$ 100 kpc.
Therefore, the larger-than-normal $EW({\rm Ly}\alpha)$ is one
of important properties of our model.

\subsection{A Possible Evolutionary Link Between LABs and Dust-enshrouded
Submm Sources}

As mentioned in section 2.2,
the central starburst region in a forming elliptical galaxy could be enshrouded
by a lot of gas with dust grains because these grain
are expected to be supplied either from Population III objects
(if any) or first massive stars in the initial starburst or both.
Therefore, elliptical galaxies at this phase may be observed as 
dust-enshrouded (or dusty) submm sources (hereafter DSSs).
Subsequent supernova explosions blow out the gas into the IGM
as a superwind several times $10^8$ years after the onset of the initial
starburst. Elliptical galaxies at this superwind phase are assumed to be
LABs in our model. We note that they are expected to be much 
fainter at submm than the DSSs because a significant part of dust grains
were already expelled out from the galaxy. In summary, the
dissipative-collapse formation of elliptical galaxies together with
the galactic wind model suggests the following evolutionary sequence.
Step I: The initial starburst occurs in the center of pregalactic gas cloud. 
Step II: This galaxy may be hidden by surrounding gas clouds for 
the first several times $10^8$ years (i.e., the DSS phase).  
Step III: The superwind blows and thus the DSS phase ceases.
The superwind leads to the formation of extended emission-line regions
around the galaxy (i.e., the LAB phase). 
This lasts for a duration of $\sim 1 \times 10^8$ years.
And, Step IV: The galaxy evolves to an ordinary elliptical galaxy
$\sim 10^9$ years after the formation.

\section{CONCLUDING REMARKS}

The origin of Ly$\alpha$ emission from high-$z$ objects may be
heterogeneous; i.e., a) ionized gas irradiated by massive stars,
b) ionized gas heated by superwinds, and c) ionized gas irradiated by 
the central engine of various types of AGNs. Such diversity is also
reported for submm-selected galaxies (i.e., DSSs) with $z > 1$
(Ivison et al. 1999). Therefore, in order to investigate the cosmic
star formation history from high-$z$ to the present day (e.g., 
Madau et al. 1996), we will have to study carefully what the observed
Ly$\alpha$ emitters at high redshift are.

\vspace {0.5cm}

We would like to thank an anonymous referee for useful suggestions 
and comments.
YS is a JSPS fellow. This work was financially supported in part by
the Ministry of Education, Science, and Culture
(Nos. 07044054, 10044052, and 10304013).

%----------------------------------------------------------------------------
%      References 
%----------------------------------------------------------------------------

%----------------------------------------------------------------------------
%     Table 1
%----------------------------------------------------------------------
\newpage

{
\small
\begin{deluxetable}{lcccccccc}
\tablewidth{6.5in}
\tablecaption{%
A summary of the three Ly$\alpha$-emitter surveys
}

\tablehead{
   \colhead{Survey\tablenotemark{a}} & 
   \colhead{Field\tablenotemark{b}} & 
   \colhead{$z_{\rm c}$\tablenotemark{c}} & 
   \colhead{$(z_{\rm min}, z_{\rm max})$\tablenotemark{d}} &
   \colhead{$V$\tablenotemark{e}} &
   \colhead{$EW_{\rm lim}({\rm Ly}\alpha)$\tablenotemark{f}} &
   \colhead{$N({\rm Ly}\alpha)$\tablenotemark{g}} &
   \colhead{$N({\rm LAB})$\tablenotemark{h}} & 
   \colhead{$n({\rm LAB})$\tablenotemark{i}} 
}
\startdata
CH98 & HDF & 3.4 & (3.41, 3.47) & 445 & 115 & 5 & 0 & 0 \nl
CH98 & SSA22 & 3.4 & (3.41, 3.47) & 445 & 90 & 7 & 0 & 0 \nl
\hline
K99 & 53W002 & 2.4 & (2.32, 2.45) & 8187 & 92 & 19 & 3\tablenotemark{j}
    & $3.66 \times 10^{-4}$ \nl
K99 & HU Aqr & 2.4 & (2.32, 2.45) & 8187 & 241 & 1 & 0 & 0 \nl
K99 & NGC 6251 & 2.4 & (2.32, 2.45) & 8187 & \nodata & 0 & 0 & 0 \nl
K99 & 53W002E & 2.55 & (2.49, 2.61) & 7417 & 291 & 1 & 0 & 0 \nl
K99 & 53W002N & 2.55 & (2.49, 2.61) & 7417 & 155 & 1 & 0 & 0 \nl
K99 & 53W002NE & 2.55 & (2.49, 2.61) & 7417 & 184 & 4 & 0 & 0 \nl
\hline
S99 & LBGS\tablenotemark{k} & 3.09 & (3.07, 3.12) & 1380 & 80 & 72 
& 2 & $1.45 \times 10^{-3}$ \nl
\enddata
\tablenotetext{a}{CH98 = Cowie \& Hu (1998), K99 = Keel et al. (1999),
     and S99 = Steidel et al. (1999).}
\tablenotetext{b}{The name of the targeted field.}
\tablenotetext{c}{The central redshift corresponding to the central
     wavelength of the narrowband filter ($\lambda_{\rm c}$).}
\tablenotetext{d}{The minimum and maximum redshift covered by the
      narrowband filter.}
\tablenotetext{e}{The co-moving volume covered by the survey in 
      units of $h^{-3}$  Mpc$^3$.}
\tablenotetext{f}{The smallest equivalent width of the Ly$\alpha$ emission
      detected in the survey in units of \AA ~ in the observed frame.}
\tablenotetext{g}{The number of Ly$\alpha$ emitters found in the survey.}
\tablenotetext{h}{The number of LABs found in the survey.}
\tablenotetext{i}{The number density of LABs found in the survey in units of
      $h^3$ Mpc$^{-3}$.}
\tablenotetext{j}{Associated with AGN.}
\tablenotetext{k}{The LBG (Lyman break galaxies) spike region.}
\end{deluxetable}
}

%----------------------------------------------------------------------------
%     Table 2
%----------------------------------------------------------------------
\newpage

{
\small
\begin{deluxetable}{lcccccc}
\tablewidth{6.5in}
\tablecaption{%
A summary of the five Ly$\alpha$ blobs
}

\tablehead{
   \colhead{Survey\tablenotemark{a}} &
   \colhead{Field\tablenotemark{b}} &
   \colhead{Name\tablenotemark{c}} &
   \colhead{$z$\tablenotemark{d}} &
   \colhead{$EW_{\rm obs}({\rm Ly}\alpha)$\tablenotemark{e}} &
   \colhead{$F({\rm Ly}\alpha)$\tablenotemark{f}} &
   \colhead{$L({\rm Ly}\alpha)$\tablenotemark{g}} 
}
\startdata
K99 & 53W002 & 53W002    & 2.390\tablenotemark{h} & 164 & 
$3.8\times10^{-16}$ & $3.9\times10^{42}$ \nl
K99 & 53W002 & Object 18 & 2.393\tablenotemark{h} & 342 & 
$1.1\times10^{-15}$ & $1.1\times10^{43}$ \nl
K99 & 53W002 & Object 19 & 2.397\tablenotemark{h} & 230 & 
$4.5\times10^{-16}$ & $4.6\times10^{42}$ \nl
\hline
S99 & LBGS & Blob 1 & 3.108 & $\sim$1500 & $1.4\times10^{-15}$ &
$2.6\times10^{43}$ \nl
S99 & LBGS & Blob 2 & 3.091 & $\sim$1500 & $1.2\times10^{-15}$ & 
$2.2\times10^{43}$ \nl
\enddata
\tablenotetext{a}{CH98 = Cowie \& Hu (1998), K99 = Keel et al. (1999),
     and S99 = Steidel et al. (1999).}
\tablenotetext{b}{The name of the targeted field.}
\tablenotetext{c}{The name of the LAB.}
\tablenotetext{d}{The observed redshift.}
\tablenotetext{e}{The observed equivalent width of the Ly$\alpha$ emission
     in units of \AA.}
\tablenotetext{f}{The observed Ly$\alpha$ flux in units of 
     ergs cm$^{-2}$ s$^{-1}$.}
\tablenotetext{g}{The Ly$\alpha$ luminosity in units of $h^{-2}$ ergs s$^{-1}$.}
\tablenotetext{h}{Taken from Pascarelle et al. (1996).}
\end{deluxetable}
}

\end{document}